\documentclass[conference,10pt,twocolumn,letter]{IEEEtran}
\IEEEoverridecommandlockouts
\usepackage{cite}
\usepackage[T1]{fontenc}
\usepackage{graphicx}
\usepackage{amssymb}
\usepackage{amsmath}
\usepackage{subfigure}
\usepackage{booktabs}
\usepackage{amsthm}
\usepackage{multirow}
\usepackage{microtype}
\usepackage{tablefootnote}
\usepackage{balance}
\usepackage{xcolor}

\usepackage{amssymb}
\usepackage{amsfonts}
\usepackage{mathrsfs}
\usepackage{xspace}
\usepackage{bm}
\usepackage{upgreek}

\newcommand{\safemath}[2]{\newcommand{#1}{\ensuremath{#2}\xspace}}

\safemath{\bma}{\mathbf{a}}
\safemath{\bmb}{\mathbf{b}}
\safemath{\bmc}{\mathbf{c}}
\safemath{\bmd}{\mathbf{d}}
\safemath{\bme}{\mathbf{e}}
\safemath{\bmf}{\mathbf{f}}
\safemath{\bmg}{\mathbf{g}}
\safemath{\bmh}{\mathbf{h}}
\safemath{\bmi}{\mathbf{i}}
\safemath{\bmj}{\mathbf{j}}
\safemath{\bmk}{\mathbf{k}}
\safemath{\bml}{\mathbf{l}}
\safemath{\bmm}{\mathbf{m}}
\safemath{\bmn}{\mathbf{n}}
\safemath{\bmo}{\mathbf{o}}
\safemath{\bmp}{\mathbf{p}}
\safemath{\bmq}{\mathbf{q}}
\safemath{\bmr}{\mathbf{r}}
\safemath{\bms}{\mathbf{s}}
\safemath{\bmt}{\mathbf{t}}
\safemath{\bmu}{\mathbf{u}}
\safemath{\bmv}{\mathbf{v}}
\safemath{\bmw}{\mathbf{w}}
\safemath{\bmx}{\mathbf{x}}
\safemath{\bmy}{\mathbf{y}}
\safemath{\bmz}{\mathbf{z}}
\safemath{\bmzero}{\mathbf{0}}
\safemath{\bmone}{\mathbf{1}}

\bmdefine{\biad}{a}
\bmdefine{\bibd}{b}
\bmdefine{\bicd}{c}
\bmdefine{\bidd}{d}
\bmdefine{\bied}{e}
\bmdefine{\bifd}{f}
\bmdefine{\bigd}{g}
\bmdefine{\bihd}{h}
\bmdefine{\biid}{i}
\bmdefine{\bijd}{j}
\bmdefine{\bikd}{k}
\bmdefine{\bild}{l}
\bmdefine{\bimd}{m}
\bmdefine{\bind}{n}
\bmdefine{\biod}{o}
\bmdefine{\bipd}{p}
\bmdefine{\biqd}{q}
\bmdefine{\bird}{r}
\bmdefine{\bisd}{s}
\bmdefine{\bitd}{t}
\bmdefine{\biud}{u}
\bmdefine{\bivd}{v}
\bmdefine{\biwd}{w}
\bmdefine{\bixd}{x}
\bmdefine{\biyd}{y}
\bmdefine{\bizd}{z}

\bmdefine{\bixid}{\xi}
\bmdefine{\bilambdad}{\lambda}
\bmdefine{\bimud}{\mu}
\bmdefine{\bithetad}{\theta}
\bmdefine{\biphid}{\phi}
\bmdefine{\bideltad}{\delta}

\safemath{\bmia}{\biad}
\safemath{\bmib}{\bibd}
\safemath{\bmic}{\bicd}
\safemath{\bmid}{\bidd}
\safemath{\bmie}{\bied}
\safemath{\bmif}{\bifd}
\safemath{\bmig}{\bigd}
\safemath{\bmih}{\bihd}
\safemath{\bmii}{\biid}
\safemath{\bmij}{\bijd}
\safemath{\bmik}{\bikd}
\safemath{\bmil}{\bild}
\safemath{\bmim}{\bimd}
\safemath{\bmin}{\bind}
\safemath{\bmio}{\biod}
\safemath{\bmip}{\bipd}
\safemath{\bmiq}{\biqd}
\safemath{\bmir}{\bird}
\safemath{\bmis}{\bisd}
\safemath{\bmit}{\bitd}
\safemath{\bmiu}{\biud}
\safemath{\bmiv}{\bivd}
\safemath{\bmiw}{\biwd}
\safemath{\bmix}{\bixd}
\safemath{\bmiy}{\biyd}
\safemath{\bmiz}{\bizd}

\safemath{\bmxi}{\bixid}
\safemath{\bmlambda}{\bilambdad}
\safemath{\bmmu}{\bimud}
\safemath{\bmtheta}{\bithetad}
\safemath{\bmphi}{\biphid}
\safemath{\bmdelta}{\bideltad}

\safemath{\bA}{\mathbf{A}}
\safemath{\bB}{\mathbf{B}}
\safemath{\bC}{\mathbf{C}}
\safemath{\bD}{\mathbf{D}}
\safemath{\bE}{\mathbf{E}}
\safemath{\bF}{\mathbf{F}}
\safemath{\bG}{\mathbf{G}}
\safemath{\bH}{\mathbf{H}}
\safemath{\bI}{\mathbf{I}}
\safemath{\bJ}{\mathbf{J}}
\safemath{\bK}{\mathbf{K}}
\safemath{\bL}{\mathbf{L}}
\safemath{\bM}{\mathbf{M}}
\safemath{\bN}{\mathbf{N}}
\safemath{\bO}{\mathbf{O}}
\safemath{\bP}{\mathbf{P}}
\safemath{\bQ}{\mathbf{Q}}
\safemath{\bR}{\mathbf{R}}
\safemath{\bS}{\mathbf{S}}
\safemath{\bT}{\mathbf{T}}
\safemath{\bU}{\mathbf{U}}
\safemath{\bV}{\mathbf{V}}
\safemath{\bW}{\mathbf{W}}
\safemath{\bX}{\mathbf{X}}
\safemath{\bY}{\mathbf{Y}}
\safemath{\bZ}{\mathbf{Z}}

\safemath{\bZero}{\mathbf{0}}
\safemath{\bOne}{\mathbf{1}}
\safemath{\bDelta}{\mathbf{\Delta}}
\safemath{\bLambda}{\mathbf{\UpLambda}}
\safemath{\bPhi}{\mathbf{\Upphi}}
\safemath{\bSigma}{\mathbf{\Upsigma}}
\safemath{\bOmega}{\mathbf{\Upomega}}
\safemath{\bTheta}{\mathbf{\Uptheta}}

\bmdefine{\biAd}{A}
\bmdefine{\biBd}{B}
\bmdefine{\biCd}{C}
\bmdefine{\biDd}{D}
\bmdefine{\biEd}{E}
\bmdefine{\biFd}{F}
\bmdefine{\biGd}{G}
\bmdefine{\biHd}{H}
\bmdefine{\biId}{I}
\bmdefine{\biJd}{J}
\bmdefine{\biKd}{K}
\bmdefine{\biLd}{L}
\bmdefine{\biMd}{M}
\bmdefine{\biOd}{N}
\bmdefine{\biPd}{O}
\bmdefine{\biQd}{P}
\bmdefine{\biRd}{R}
\bmdefine{\biSd}{S}
\bmdefine{\biTd}{T}
\bmdefine{\biUd}{U}
\bmdefine{\biVd}{V}
\bmdefine{\biWd}{W}
\bmdefine{\biXd}{X}
\bmdefine{\biYd}{Y}
\bmdefine{\biZd}{Z}

\bmdefine{\biDelta}{\Delta}
\bmdefine{\biLambda}{\Lambda}
\bmdefine{\biPhi}{\Phi}
\bmdefine{\biSigma}{\Sigma}
\bmdefine{\biOmega}{\Omega}
\bmdefine{\biTheta}{\Theta}

\safemath{\bimA}{\biAd}
\safemath{\bimB}{\biBd}
\safemath{\bimC}{\biCd}
\safemath{\bimD}{\biDd}
\safemath{\bimE}{\biEd}
\safemath{\bimF}{\biFd}
\safemath{\bimG}{\biGd}
\safemath{\bimH}{\biHd}
\safemath{\bimI}{\biId}
\safemath{\bimJ}{\biJd}
\safemath{\bimK}{\biKd}
\safemath{\bimL}{\biLd}
\safemath{\bimM}{\biMd}
\safemath{\bimN}{\biNd}
\safemath{\bimO}{\biOd}
\safemath{\bimP}{\biPd}
\safemath{\bimQ}{\biQd}
\safemath{\bimR}{\biRd}
\safemath{\bimS}{\biSd}
\safemath{\bimT}{\biTd}
\safemath{\bimU}{\biUd}
\safemath{\bimV}{\biVd}
\safemath{\bimW}{\biWd}
\safemath{\bimX}{\biXd}
\safemath{\bimY}{\biYd}
\safemath{\bimZ}{\biZd}

\safemath{\bimDelta}{\biDelta}
\safemath{\bimLambda}{\biLambda}
\safemath{\bimPhi}{\biPhi}
\safemath{\bimSigma}{\biSigma}
\safemath{\bimOmega}{\biOmega}
\safemath{\bimTheta}{\biTheta}

\safemath{\setA}{\mathcal{A}}
\safemath{\setB}{\mathcal{B}}
\safemath{\setC}{\mathcal{C}}
\safemath{\setD}{\mathcal{D}}
\safemath{\setE}{\mathcal{E}}
\safemath{\setF}{\mathcal{F}}
\safemath{\setG}{\mathcal{G}}
\safemath{\setH}{\mathcal{H}}
\safemath{\setI}{\mathcal{I}}
\safemath{\setJ}{\mathcal{J}}
\safemath{\setK}{\mathcal{K}}
\safemath{\setL}{\mathcal{L}}
\safemath{\setM}{\mathcal{M}}
\safemath{\setN}{\mathcal{N}}
\safemath{\setO}{\mathcal{O}}
\safemath{\setP}{\mathcal{P}}
\safemath{\setQ}{\mathcal{Q}}
\safemath{\setR}{\mathcal{R}}
\safemath{\setS}{\mathcal{S}}
\safemath{\setT}{\mathcal{T}}
\safemath{\setU}{\mathcal{U}}
\safemath{\setV}{\mathcal{V}}
\safemath{\setW}{\mathcal{W}}
\safemath{\setX}{\mathcal{X}}
\safemath{\setY}{\mathcal{Y}}
\safemath{\setZ}{\mathcal{Z}}
\safemath{\emptySet}{\varnothing}

\safemath{\colA}{\mathscr{A}}
\safemath{\colB}{\mathscr{B}}
\safemath{\colC}{\mathscr{C}}
\safemath{\colD}{\mathscr{D}}
\safemath{\colE}{\mathscr{E}}
\safemath{\colF}{\mathscr{F}}
\safemath{\colG}{\mathscr{G}}
\safemath{\colH}{\mathscr{H}}
\safemath{\colI}{\mathscr{I}}
\safemath{\colJ}{\mathscr{J}}
\safemath{\colK}{\mathscr{K}}
\safemath{\colL}{\mathscr{L}}
\safemath{\colM}{\mathscr{M}}
\safemath{\colN}{\mathscr{N}}
\safemath{\colO}{\mathscr{O}}
\safemath{\colP}{\mathscr{P}}
\safemath{\colQ}{\mathscr{Q}}
\safemath{\colR}{\mathscr{R}}
\safemath{\colS}{\mathscr{S}}
\safemath{\colT}{\mathscr{T}}
\safemath{\colU}{\mathscr{U}}
\safemath{\colV}{\mathscr{V}}
\safemath{\colW}{\mathscr{W}}
\safemath{\colX}{\mathscr{X}}
\safemath{\colY}{\mathscr{Y}}
\safemath{\colZ}{\mathscr{Z}}

\safemath{\opA}{\mathbb{A}}
\safemath{\opB}{\mathbb{B}}
\safemath{\opC}{\mathbb{C}}
\safemath{\opD}{\mathbb{D}}
\safemath{\opE}{\mathbb{E}}
\safemath{\opF}{\mathbb{F}}
\safemath{\opG}{\mathbb{G}}
\safemath{\opH}{\mathbb{H}}
\safemath{\opI}{\mathbb{I}}
\safemath{\opJ}{\mathbb{J}}
\safemath{\opK}{\mathbb{K}}
\safemath{\opL}{\mathbb{L}}
\safemath{\opM}{\mathbb{M}}
\safemath{\opN}{\mathbb{N}}
\safemath{\opO}{\mathbb{O}}
\safemath{\opP}{\mathbb{P}}
\safemath{\opQ}{\mathbb{Q}}
\safemath{\opR}{\mathbb{R}}
\safemath{\opS}{\mathbb{S}}
\safemath{\opT}{\mathbb{T}}
\safemath{\opU}{\mathbb{U}}
\safemath{\opV}{\mathbb{V}}
\safemath{\opW}{\mathbb{W}}
\safemath{\opX}{\mathbb{X}}
\safemath{\opY}{\mathbb{Y}}
\safemath{\opZ}{\mathbb{Z}}
\safemath{\opZero}{\mathbb{O}}
\safemath{\identityop}{\opI}

\safemath{\veca}{\bma}
\safemath{\vecb}{\bmb}
\safemath{\vecc}{\bmc}
\safemath{\vecd}{\bmd}
\safemath{\vece}{\bme}
\safemath{\vecf}{\bmf}
\safemath{\vecg}{\bmg}
\safemath{\vech}{\bmh}
\safemath{\veci}{\bmi}
\safemath{\vecj}{\bmj}
\safemath{\veck}{\bmk}
\safemath{\vecl}{\bml}
\safemath{\vecm}{\bmm}
\safemath{\vecn}{\bmn}
\safemath{\veco}{\bmo}
\safemath{\vecp}{\bmp}
\safemath{\vecq}{\bmq}
\safemath{\vecr}{\bmr}
\safemath{\vecs}{\bms}
\safemath{\vect}{\bmt}
\safemath{\vecu}{\bmu}
\safemath{\vecv}{\bmv}
\safemath{\vecw}{\bmw}
\safemath{\vecx}{\bmx}
\safemath{\vecy}{\bmy}
\safemath{\vecz}{\bmz}

\safemath{\veczero}{\bmzero}
\safemath{\vecone}{\bmone}
\safemath{\vecxi}{\bmxi}
\safemath{\veclambda}{\bmlambda}
\safemath{\vecmu}{\bmmu}
\safemath{\vectheta}{\bmtheta}
\safemath{\vecphi}{\bmphi}
\safemath{\vecdelta}{\bmdelta}

\safemath{\matA}{\bA}
\safemath{\matB}{\bB}
\safemath{\matC}{\bC}
\safemath{\matD}{\bD}
\safemath{\matE}{\bE}
\safemath{\matF}{\bF}
\safemath{\matG}{\bG}
\safemath{\matH}{\bH}
\safemath{\matI}{\bI}
\safemath{\matJ}{\bJ}
\safemath{\matK}{\bK}
\safemath{\matL}{\bL}
\safemath{\matM}{\bM}
\safemath{\matN}{\bN}
\safemath{\matO}{\bO}
\safemath{\matP}{\bP}
\safemath{\matQ}{\bQ}
\safemath{\matR}{\bR}
\safemath{\matS}{\bS}
\safemath{\matT}{\bT}
\safemath{\matU}{\bU}
\safemath{\matV}{\bV}
\safemath{\matW}{\bW}
\safemath{\matX}{\bX}
\safemath{\matY}{\bY}
\safemath{\matZ}{\bZ}
\safemath{\matzero}{\bmzero}

\safemath{\matDelta}{\bDelta}
\safemath{\matLambda}{\bLambda}
\safemath{\matPhi}{\bPhi}
\safemath{\matSigma}{\bSigma}
\safemath{\matOmega}{\bOmega}
\safemath{\matTheta}{\bTheta}

\safemath{\matidentity}{\matI}
\safemath{\matone}{\matO}

\safemath{\rnda}{A}
\safemath{\rndb}{B}
\safemath{\rndc}{C}
\safemath{\rndd}{D}
\safemath{\rnde}{E}
\safemath{\rndf}{F}
\safemath{\rndg}{G}
\safemath{\rndh}{H}
\safemath{\rndi}{I}
\safemath{\rndj}{J}
\safemath{\rndk}{K}
\safemath{\rndl}{L}
\safemath{\rndm}{M}
\safemath{\rndn}{N}
\safemath{\rndo}{O}
\safemath{\rndp}{P}
\safemath{\rndq}{Q}
\safemath{\rndr}{R}
\safemath{\rnds}{S}
\safemath{\rndt}{T}
\safemath{\rndu}{U}
\safemath{\rndv}{V}
\safemath{\rndw}{W}
\safemath{\rndx}{X}
\safemath{\rndy}{Y}
\safemath{\rndz}{Z}

\safemath{\rveca}{\bimA}
\safemath{\rvecb}{\bimB}
\safemath{\rvecc}{\bimC}
\safemath{\rvecd}{\bimD}
\safemath{\rvece}{\bimE}
\safemath{\rvecf}{\bimF}
\safemath{\rvecg}{\bimG}
\safemath{\rvech}{\bimH}
\safemath{\rveci}{\bimI}
\safemath{\rvecj}{\bimJ}
\safemath{\rveck}{\bimK}
\safemath{\rvecl}{\bimL}
\safemath{\rvecm}{\bimM}
\safemath{\rvecn}{\bimN}
\safemath{\rveco}{\bomO}
\safemath{\rvecp}{\bimP}
\safemath{\rvecq}{\bimQ}
\safemath{\rvecr}{\bimR}
\safemath{\rvecs}{\bimS}
\safemath{\rvect}{\bimT}
\safemath{\rvecu}{\bimU}
\safemath{\rvecv}{\bimV}
\safemath{\rvecw}{\bimW}
\safemath{\rvecx}{\bimX}
\safemath{\rvecy}{\bimY}
\safemath{\rvecz}{\bimZ}

\safemath{\rvecxi}{\bmxi}
\safemath{\rveclambda}{\bmlambda}
\safemath{\rvecmu}{\bmmu}
\safemath{\rvectheta}{\bmtheta}
\safemath{\rvecphi}{\bmphi}

\safemath{\rmatA}{\bimA}
\safemath{\rmatB}{\bimB}
\safemath{\rmatC}{\bimC}
\safemath{\rmatD}{\bimD}
\safemath{\rmatE}{\bimE}
\safemath{\rmatF}{\bimF}
\safemath{\rmatG}{\bimG}
\safemath{\rmatH}{\bimH}
\safemath{\rmatI}{\bimI}
\safemath{\rmatJ}{\bimJ}
\safemath{\rmatK}{\bimK}
\safemath{\rmatL}{\bimL}
\safemath{\rmatM}{\bimM}
\safemath{\rmatN}{\bimN}
\safemath{\rmatO}{\bimO}
\safemath{\rmatP}{\bimP}
\safemath{\rmatQ}{\bimQ}
\safemath{\rmatR}{\bimR}
\safemath{\rmatS}{\bimS}
\safemath{\rmatT}{\bimT}
\safemath{\rmatU}{\bimU}
\safemath{\rmatV}{\bimV}
\safemath{\rmatW}{\bimW}
\safemath{\rmatX}{\bimX}
\safemath{\rmatY}{\bimY}
\safemath{\rmatZ}{\bimZ}

\safemath{\rmatDelta}{\bimDelta}
\safemath{\rmatLambda}{\bimLambda}
\safemath{\rmatPhi}{\bimPhi}
\safemath{\rmatSigma}{\bimSigma}
\safemath{\rmatOmega}{\bimOmega}
\safemath{\rmatTheta}{\bimTheta}

\usepackage{amssymb}
\usepackage{amsfonts}
\usepackage{mathrsfs}
\usepackage{xspace}
\usepackage{bm}
\usepackage{fancyref}
\usepackage{textcomp}

\usepackage{multirow}
\usepackage{stmaryrd}

\newenvironment{textbmatrix}{	\setlength{\arraycolsep}{2.5pt}%
								\big[\begin{matrix}}{\end{matrix}\big]%
								\raisebox{0.08ex}{\vphantom{M}}}

\def\be{\begin{equation}}
\def\ee{\end{equation}}
\def\een{\nonumber \end{equation}}
\def\mat{\begin{bmatrix}}
\def\emat{\end{bmatrix}}
\def\btm{\begin{textbmatrix}}
\def\etm{\end{textbmatrix}}

\def\ba#1\ea{\begin{align}#1\end{align}}
\def\bas#1\eas{\begin{align*}#1\end{align*}}
\def\bs#1\es{\begin{split}#1\end{split}}
\def\bg#1\eg{\begin{gather}#1\end{gather}}
\def\bml#1\eml{\begin{multline}#1\end{multline}}
\def\bi#1\ei{\begin{itemize}#1\end{itemize}}

\newcommand{\Ex}[2]{\ensuremath{\Exop_{#1}\lefto[#2\right]}} 	

\safemath{\dirac}{\delta}					
\safemath{\krond}{\dirac}					

\safemath{\upto}{\uparrow}
\safemath{\downto}{\downarrow}
\safemath{\iu}{j}							
\safemath{\ev}{\lambda}						
\safemath{\hilseqspace}{l^{2}}				
\newcommand{\banachfunspace}[1]{\setL^{#1}}	
\safemath{\hilfunspace}{\banachfunspace{2}}	
\newcommand{\floor}[1]{\lfloor #1 \rfloor}

\safemath{\SNR}{\textit{SNR}} 				
\safemath{\PAR}{\textit{PAR}} 				
\safemath{\No}{N_0}							
\safemath{\Es}{E_s}							
\safemath{\Eb}{E_b}							
\safemath{\EbNo}{\frac{\Eb}{\No}}
\safemath{\EsNo}{\frac{\Es}{\No}}

\DeclareMathOperator{\CHop}{\ensuremath{\opH}} 
\safemath{\tvir}{\rndh_{\CHop}}				
\safemath{\tvtf}{\rndl_{\CHop}}				
\safemath{\spf}{\rnds_{\CHop}}				
\safemath{\bff}{H_{\CHop}}					

\safemath{\ircf}{r_{h}}						
\safemath{\tftvcf}{r_{s}}					
\safemath{\tfcf}{r_{l}}						
\safemath{\bfcf}{r_{H}}						

\safemath{\tcorr}{c_h}						
\safemath{\scf}{c_{s}}						
\safemath{\tfcorr}{c_{l}}					
\safemath{\fcorr}{c_{H}}						

\safemath{\mi}{I}							
\safemath{\capacity}{C}						

\safemath{\normal}{\mathcal{N}}			
\safemath{\jpg}{\mathcal{CN}}			
\safemath{\mchain}{\leftrightarrow}		

\safemath{\dB}{\,\mathrm{dB}}
\safemath{\dBm}{\,\mathrm{dBm}}
\safemath{\Hz}{\,\mathrm{Hz}}
\safemath{\kHz}{\,\mathrm{kHz}}
\safemath{\MHz}{\,\mathrm{MHz}}
\safemath{\GHz}{\,\mathrm{GHz}}
\safemath{\s}{\,\mathrm{s}}
\safemath{\ms}{\,\mathrm{ms}}
\safemath{\mus}{\,\mathrm{\text{\textmu}s}}
\safemath{\ns}{\,\mathrm{ns}}
\safemath{\ps}{\,\mathrm{ps}}
\safemath{\meter}{\,\mathrm{m}}
\safemath{\mm}{\,\mathrm{mm}}
\safemath{\cm}{\,\mathrm{cm}}
\safemath{\m}{\,\mathrm{m}}
\safemath{\W}{\,\mathrm{W}}
\safemath{\mW}{\, \mathrm{mW}}
\safemath{\J}{\,\mathrm{J}}
\safemath{\K}{\,\mathrm{K}}
\safemath{\bit}{\,\mathrm{bit}}
\safemath{\nat}{\,\mathrm{nat}}

\safemath{\define}{\triangleq}			

\safemath{\equivalent}{\sim}
\safemath{\distas}{\sim}					
\safemath{\sdiff}{\Delta}				

\safemath{\reals}{\mathbb{R}}
\safemath{\positivereals}{\reals_{+}}
\safemath{\integers}{\mathbb{Z}}
\safemath{\posint}{\integers_{+}}
\safemath{\naturals}{\mathbb{N}}
\safemath{\posnaturals}{\naturals_{+}}
\safemath{\complexset}{\mathbb{C}}
\safemath{\rationals}{\mathbb{Q}}

\newcommand*{\fancyrefapplabelprefix}{app}		
\newcommand*{\fancyrefthmlabelprefix}{thm}		
\newcommand*{\fancyreflemlabelprefix}{lem}		
\newcommand*{\fancyrefcorlabelprefix}{cor}		
\newcommand*{\fancyrefdeflabelprefix}{def}		
\newcommand*{\fancyrefproplabelprefix}{prop}		
\newcommand*{\fancyrefexmpllabelprefix}{exmpl}
\newcommand*{\fancyrefalglabelprefix}{alg}		
\newcommand*{\fancyreftbllabelprefix}{tbl}		

\frefformat{vario}{\fancyrefseclabelprefix}{Section~#1}
\frefformat{vario}{\fancyrefthmlabelprefix}{Theorem~#1}
\frefformat{vario}{\fancyreftbllabelprefix}{Table~#1}
\frefformat{vario}{\fancyreflemlabelprefix}{Lemma~#1}
\frefformat{vario}{\fancyrefcorlabelprefix}{Corollary~#1}
\frefformat{vario}{\fancyrefdeflabelprefix}{Definition~#1}
\frefformat{vario}{\fancyreffiglabelprefix}{Figure~#1}
\frefformat{vario}{\fancyrefapplabelprefix}{Appendix~#1}
\frefformat{vario}{\fancyrefeqlabelprefix}{(#1)}
\frefformat{vario}{\fancyrefproplabelprefix}{Proposition~#1}
\frefformat{vario}{\fancyrefexmpllabelprefix}{Example~#1}
\frefformat{vario}{\fancyrefalglabelprefix}{Algorithm~#1}

\safemath{\dictab}{[\,\dicta\,\,\dictb\,]}

\safemath{\ysig}{\bmy}
\safemath{\ysighat}{\hat{\ysig}}
\safemath{\ysigdim}{M}
\safemath{\xsig}{\bmx}
\safemath{\xsigdim}{N}
\safemath{\nx}{n_x}
\safemath{\zsig}{\bmz}
\safemath{\zsigdim}{\ysigdim}
\safemath{\rsig}{\bmr}
\safemath{\Adict}{\bA}
\safemath{\Adicttilde}{\widetilde{\Adict}}
\safemath{\Adictdim}{\outputdim\times\xsigdim}
\safemath{\avec}{\bma}
\safemath{\avectilde}{\tilde{\avec}}
\safemath{\Bdict}{\bB}
\safemath{\Bdicttilde}{\widetilde{\Bdict}}
\safemath{\Cdict}{\bC}
\safemath{\cvec}{\bmc}
\safemath{\Ddict}{\bD}
\safemath{\Ddictdim}{\ysigdim\times\xsigdim}
\safemath{\dvec}{\bmd}
\safemath{\Ddicttilde}{\widetilde{\bD}}
\safemath{\Bonb}{\bB}
\safemath{\bvec}{\bmb}
\safemath{\Bonbdim}{\ysigdim\times\ysigdim}
\safemath{\noise}{\bmn}
\safemath{\noisedim}{\ysigim}
\safemath{\err}{\bme}
\safemath{\errdim}{\ysigdim}
\safemath{\errset}{\setE}
\safemath{\nerr}{n_e}
\safemath{\delop}{\bP_\errset}
\safemath{\delopc}{\bP_{{\errset}^c}}

\safemath{\cplxi}{\imath}
\safemath{\cplxj}{\jmath}

\safemath{\dict}{\matD}
\safemath{\inputdim}{N}		
\safemath{\outputdim}{M}		
\safemath{\sparsity}{S}	
\safemath{\inputdimA}{{N_a}}	
\safemath{\inputdimB}{{N_b}}	
\safemath{\elemA}{{n_a}}	
\safemath{\elemB}{{n_b}}	
\safemath{\resA}{\matR_a}	
\safemath{\resB}{\matR_b}	
\safemath{\subD}{\matS} 
\safemath{\subA}{\matS_a} 
\safemath{\subB}{\matS_b} 
\safemath{\dicta}{\matA} 	
\safemath{\dictb}{\matB} 	
\safemath{\hollowS}{H}
\safemath{\hollowA}{H_a}
\safemath{\hollowB}{H_b}
\safemath{\cross}{Z}
\safemath{\coh}{\mu_d}			
\safemath{\coha}{\mu_a}			
\safemath{\cohb}{\mu_b}			
\safemath{\mubs}{\nu}	
\safemath{\cohm}{\mu_m} 
\safemath{\dictset}{\setD}	
\safemath{\dictsetp}{\dictset(\coh,\coha,\cohb)}	
\safemath{\dictsetgen}{\dictset_\text{gen}}
\safemath{\dictsetgenp}{\dictsetgen(\coh)}
\safemath{\dictsetonb}{\dictset_\text{onb}}
\safemath{\dictsetonbp}{\dictsetonb(\coh)}

\safemath{\leftside}{U}
\safemath{\rightsideA}{R_a}
\safemath{\rightsideB}{R_b}

\safemath{\indexS}{\setI_S} 

\safemath{\na}{n_a}			
\safemath{\nb}{n_b}			
\safemath{\coeffa}{p_i}	
\safemath{\coeffb}{q_j}	
\safemath{\seta}{\setP}		
\safemath{\setb}{\setQ}     
\safemath{\setw}{\setW}	
\safemath{\setz}{\setZ}	
\safemath{\cola}{\veca}		
\safemath{\colb}{\vecb}		
\safemath{\cold}{\vecd}		
\safemath{\inputvec}{\vecx} 	
\safemath{\error}{\vece}	
\safemath{\noiseout}{\vecz} 	
\safemath{\inputvecel}{x}
\safemath{\inputveca}{\vecx_a}
\safemath{\inputvecb}{\vecx_b}
\safemath{\outputvec}{\vecy}	
\safemath{\lambdamin}{\lambda_{\mathrm{min}}}

\safemath{\elltwo}{\ell_2}
\safemath{\ellone}{\ell_1}
\safemath{\ellzero}{\ell_0}
\safemath{\ellinf}{\ell_\infty}
\safemath{\ellinftilde}{\ell_{\widetilde\infty}}
\safemath{\licard}{Z(\coh,\coha,\cohb)}
\safemath{\xsol}{\hat{x}}
\safemath{\xbord}{x_b}		
\safemath{\xstat}{x_s}		
\safemath{\xstatLone}{\tilde{x}_s}
\safemath{\order}{\mathcal{O}} 
\safemath{\scales}{\Theta} 
\safemath{\ones}{\mathbf{1}} 
\safemath{\zeroes}{\mathbf{0}} 
\safemath{\thlone}{\kappa(\coh,\cohb)} 
\safemath{\constoneA}{\delta} 
\safemath{\constoneB}{\epsilon} 
\safemath{\nlarge}{L}				   
\safemath{\sumlarge}{S_\nlarge}
\safemath{\maxlarger}{P_\nlarge}	   
\safemath{\Pzero}{\textrm{P0}}	
\safemath{\Pone}{\textrm{P1}}
\safemath{\vecfir}{\vecw}			 
\safemath{\vecsec}{\vecz}
\safemath{\elvecfir}{w}              
\safemath{\elvecsec}{z}				 
\safemath{\nlargefir}{n}
\safemath{\normout}{\gamma}
\safemath{\auxfun}{h}
\safemath{\supp}{\textrm{supp}}

\safemath{\indexa}{\ell}
\safemath{\indexb}{r}
\safemath{\indexc}{i}
\safemath{\indexd}{j}

\safemath{\project}{P}

\allowdisplaybreaks

\safemath{\LAMA}{\textrm{LAMA}}
\safemath{\MRT}{\textrm{MRT}}
\safemath{\betamax}{\beta^\text{max}_\setO}
\safemath{\betamaxno}{\beta^\text{max}}
\safemath{\betamin}{\beta^\text{min}_\setO}
\safemath{\betaminno}{\beta^\text{min}}

\safemath{\Nomin}{\No^\textnormal{min}(\beta)}
\safemath{\Nominnobeta}{\No^\text{min}}
\safemath{\Nomax}{\No^\textnormal{max}(\beta)}
\safemath{\Nomaxnobeta}{\No^\textnormal{max}}
\safemath{\EX}{E_\textnormal{x}}
\safemath{\EXP}{\EX^\textnormal{p}}
\safemath{\Eo}{E_0}

\safemath{\tmax}{{t_\textnormal{max}}}
\safemath{\MAP}{\textrm{MAP}}
\safemath{\IO}{\textrm{IO}}
\safemath{\JO}{\textrm{JO}}
\safemath{\Nopost}{N_{0}^\textnormal{post}}
\safemath{\MT}{U}
\safemath{\MR}{B}
\safemath{\Tran}{\textnormal{T}}
\safemath{\Herm}{\textnormal{H}}
\safemath{\row}{\textnormal{r}}
\safemath{\col}{\textnormal{c}}

\safemath{\NT}{N_\textnormal{T}}
\safemath{\DSNR}{\delta \textnormal{SNR}}
\safemath{\betaMOR}{\beta^{\star}}

\begin{document}
	
\title{Resolution-Adaptive All-Digital Spatial Equalization \\ for mmWave Massive MU-MIMO}%
\author{\IEEEauthorblockN{Oscar Casta\~neda$^1$, Seyed Hadi Mirfarshbafan$^1$, Shahaboddin Ghajari$^2$, \\ Alyosha Molnar$^2$, Sven Jacobsson$^3$, Giuseppe Durisi$^4$, and Christoph Studer$^1$} \\[-0.0cm]
\IEEEauthorblockA{\textit{$^1$Department of Information Technology
and Electrical Engineering, ETH Z\"urich, Z\"urich, Switzerland} \\
\textit{$^2$School of Electrical and Computer Engineering, Cornell University, Ithaca, NY, USA} \\
\textit{$^3$Ericsson Research, Gothenburg, Sweden} \\
\textit{$^4$Department of Electrical Engineering, Chalmers University of Technology, Gothenburg, Sweden}}
\thanks{The work of OC, SHM, AM, and CS was supported in part by  ComSenTer, one of six centers in JUMP, a SRC program sponsored by DARPA. The work of CS was also supported by an ETH Research Grant and by the US NSF under grants CNS-1717559, CNS-1955997, and ECCS-1824379.}\thanks{Contact author: O. Casta\~neda; e-mail: caoscar@ethz.ch}\\[-0.4cm]
}

\maketitle


\begin{abstract}
All-digital basestation (BS) architectures for millimeter-wave (mmWave) massive multi-user multiple-input multiple-output (MU-MIMO), which equip each radio-frequency chain with dedicated data converters, have advantages in spectral efficiency, flexibility, and baseband-processing simplicity over hybrid analog-digital solutions.
For all-digital architectures to be competitive with hybrid solutions in terms of power consumption, novel signal-processing methods and baseband architectures are necessary. 
In this paper, we demonstrate that adapting the resolution of the analog-to-digital converters (ADCs) and spatial equalizer of an all-digital system to the communication scenario (e.g., the number of users, modulation scheme, and propagation conditions) enables orders-of-magnitude power savings for realistic mmWave channels.
For example, for a 256-BS-antenna 16-user system supporting 1\,GHz bandwidth, a traditional baseline architecture designed for a 64-user worst-case scenario would consume 23\,W in 28\,nm CMOS for the ADC array and the spatial equalizer, whereas a resolution-adaptive architecture is able to reduce the power consumption by  6.7$\boldsymbol\times$.
\end{abstract}


\section{Introduction} 
\label{sec:intro}

Millimeter-wave (mmWave) communication~\cite{rappaport13a,rappaport17} offers vast amounts of unused spectrum and is considered a key technology component of fifth-generation (5G) and beyond 5G wireless  systems. A major challenge of mmWave communication is the high path loss~\cite{rappaport15a}, which can be mitigated with massive multiple-input multiple-output (MIMO)~\cite{larsson14a}. 
Massive multi-user (MU) MIMO is able to compensate for the high path loss via fine-grained beamforming while supporting concurrent communication with multiple user equipments~(UEs) in the same frequency band. 
A practical realization of mmWave massive MU-MIMO basestations (BSs), however, faces serious implementation challenges that are caused by the large number of BS antennas and the large communication bandwidth. 

Two prominent solutions that lower the cost and power consumption of mmWave massive MU-MIMO systems are (i) hybrid digital-analog architectures  \cite{sohrabi16, heath-jr.15a} in which the number of radio-frequency (RF) chains is less than the number of antennas and (ii) all-digital architectures that rely on nonlinear RF chains and low-resolution data converters~\cite{abdelghany18, jacobsson18d, Xu19}. In this paper, we focus on all-digital architectures that are able to exploit the full potential of massive MU-MIMO and offer higher spectral efficiency with comparable power consumption to hybrid architectures~\cite{yan2019performance, panagiotis20}. However, in order to keep the system costs and power consumption of all-digital architectures within practical limits, it is indispensable to rely on low-resolution data converters~\cite{jacobsson17b} as well as on low-resolution digital baseband processing methods~\cite{castaneda19fame}.

\subsection{Contributions}
In this paper, we show that for all-digital massive MU-MIMO systems with $B$ BS antennas and $U$ UEs, the system's load factor defined as $\beta\define U/B$, as well as other specifics of the communication system (e.g., the modulation and channel conditions) do not only determine the  spectral efficiency~\cite{Jeon15} and robustness against RF impairments~\cite{gustavsson14a, bjornson15c, jacobsson17b}, but also open up new means to optimize the power consumption of all-digital spatial equalization architectures. 
More specifically, if the number of active UEs is low, then it is possible to reduce the resolution of the analog-to-digital converters (ADCs) and the resolution of finite-alphabet equalizers, as well as the number of active BS antennas, without noticeably degrading the system's error-rate performance.
We demonstrate that such \emph{resolution-adaptive} all-digital massive MU-MIMO BS architectures are key to enabling up to two orders-of-magnitude power savings in the ADC array and the all-digital spatial equalizer. 

Previous work~\cite{liang15a,choi17a,verenzuela21a} focused on adapting the resolution of each individual ADC to the channel state.
In contrast, we (i) adapt the resolutions of both the ADC array \emph{and} the spatial equalizer to the instantaneous communication scenario and (ii) study their individual impact on power consumption.

\subsection{Notation}
Boldface uppercase and lowercase letters represent matrices and column vectors, respectively.
For a matrix $\bA$, the conjugate transpose is $\bA^\Herm$, the $k$th column is $\bma_k$, and the entry on the $k$th row and $\ell$th column is $A_{k,\ell}$.
The $N\times N$ identity matrix is $\bI_N$.
For a vector $\bma$, the $k$th entry is $a_k$, the $\ell_2$-norm is $\|\bma\|_2$, the real part is $\Re\{\bma\}$ and the imaginary parts is $\Im\{\bma\}$.
The $\ell_\infty$-norm and $\ell_{\widetilde\infty}$-norm are defined as $\|\bma\|_\infty \define \max_{k} |a_k|$ and $\|\bma\|_{\widetilde\infty} \define \max\{\|\Re\{\bma\}\|_\infty,\|\Im\{\bma\}\|_\infty\}$.
Expectation with respect to the random vector $\bmx$ is denoted by \Ex{\bmx}{\cdot}.


\section{System Model}
\label{sec:system}

\begin{figure}[tp]
\centering
\includegraphics[width=0.9\columnwidth]{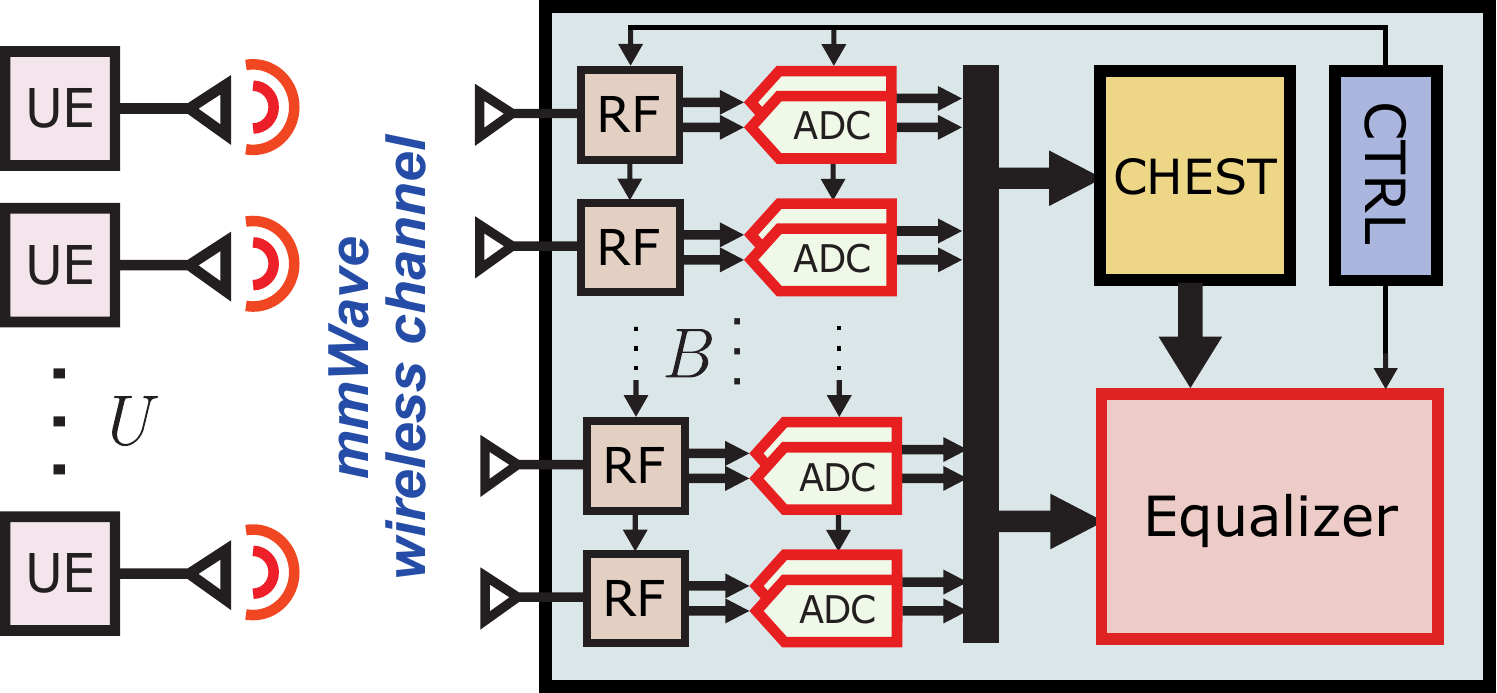}
\caption{System overview of a resolution-adaptive all-digital mmWave massive MU-MIMO BS equipped with $B$ antennas, a pair of ADCs per RF chain, and a digital equalizer. A controller (CTRL) adapts the number of active BS antennas $B'$ as well as the resolution of the ADCs and equalizer based on the communication scenario in order to minimize the system's power consumption.}
\label{fig:system_overview}
\end{figure}

We consider the uplink of a mmWave massive MU-MIMO system in which $U$ single-antenna UEs transmit to a resolution-adaptive all-digital BS with $B$ antennas as illustrated in \fref{fig:system_overview}.
At the BS, we model the nonlinear distortion artifacts caused by finite-resolution ADCs and finite-alphabet spatial equalization. Our goal is to characterize the performance/power trade-offs of such a resolution-adaptive all-digital equalization architecture. 

\subsection{System Model}
We model wireless transmission with the following frequency-flat input-output relation: $\bmy = \bH\bms + \bmn$.
Here, $\bmy\in\complexset^B$ is the (unquantized) receive vector at the BS, $\bH\in\complexset^{B\times U}$ is the MIMO channel matrix, $\bms\in\setS^U$ is the transmit vector in which each entry corresponds to the per-UE transmit symbol taken from a constellation set $\setS$  (e.g., $16$-QAM), and $\bmn\in\complexset^B$ models i.i.d.\ circularly-symmetric complex Gaussian noise with a per-entry variance of $\No$.
In what follows, we assume that the UE transmit symbols $s_u$, $u=1,\dots,U$, are independent and zero mean, each with a variance of $\Es\sigma^2_u$.
We further assume $\pm3\,$dB power control across UEs so that $\max_{u}\{\sigma^2_u\|\bmh_u\|^2\}/\min_{u}\{\sigma^2_u\|\bmh_u\|^2\}=4$ \cite{castaneda19fame}.
For this model, the average receive signal-to-noise ratio (SNR)~is 
\begin{equation}
\textit{SNR} \define \frac{\Ex{\bms}{\|\bH\bms\|_2^2}}{\Ex{\bmn}{\|\bmn\|_2^2}} = \frac{\Es \sum_{u=1}^{U}{\sigma^2_u\|\bmh_u\|_2^2}}{\No B},
\end{equation}
where $\bmh_u\in\complexset^{B}$, $u=1,\dots,U$, is the channel for the $u$th UE.
We further consider that the BS can control the number of \emph{active} antennas $B'\leq B$, so that only a subset $\hat{\bmy}\in\complexset^{B'}$ of the signals in $\bmy$ are further processed.
In our case, the BS will select the $B'$ contiguous antennas at the center of the array.

\subsection{Analog-to-Digital Conversion}
In order to reduce the power consumption of all-digital BS architectures, practical systems will have to rely on low-resolution data converters~\cite{abdelghany18, jacobsson18d, Xu19}.
To take into account the quantization artifacts caused by such low-resolution ADCs, we assume that the receive (active) signal $\hat{\bmy}$ passes through $2B'$ quantizers, $B'$ for the in-phase and quadrature (I/Q) components, resulting in the quantized receive vector $\bmz$, where $\bmz=Q_q(\Re\{\hat{\bmy}\})+jQ_q(\Im\{\hat{\bmy}\})$.
Here, the quantization function $Q_q$ is applied entry-wise to its argument and models a $q$-bit uniform midrise quantizer with step size $\Delta$ as, e.g., in \cite{jacobsson18d} 
\begin{equation} \label{eq:quantizer}
Q_q(y)=
    \begin{cases}
      \Delta\floor{\frac{y}{\Delta}}+\frac{\Delta}{2}, & \text{if}\ |y|<\Delta 2^{q-1}\\
      \frac{\Delta}{2}(2^q-1)\frac{y}{|y|}, & \text{if}\ |y|>\Delta 2^{q-1}.
    \end{cases}
\end{equation}
We use the step size $\Delta$ that minimizes the mean-squared error (MSE) between the quantizer's input~$y$ and output~$Q_q(y)$ assuming that the ADCs' input is a circularly-symmetric complex Gaussian random variable~\cite{max60a} with variance equal to
\begin{align}
\sigma^2_\text{ADC} = \No + \Es \max_{b\in \{1,\ldots,B'\}} \textstyle \sum_{u=1}^{U} \sigma^2_u |H_{b,u}|^2, 
\end{align}
which is the maximum variance across all active receive antennas.
In practice, the step size would be adjusted by an automatic gain control circuit. 

\subsection{Finite-Alphabet Equalization}
Given the extremely large bandwidths available at mmWave frequencies, linear equalization is preferred to keep implementation complexity within reasonable bounds.
A linear equalizer generates estimates $\hat{\bms}$ of the transmit vector $\bms$ according to $\hat{\bms}=\bW^\Herm\bmz$, where $\bW^\Herm\in\complexset^{U\times B'}$ is a suitably-designed equalization matrix.
The equalization matrix $\bW^\Herm$ is typically designed to minimize the post-equalization MSE defined as $\textit{MSE}\define\Ex{\bms,\bmn}{\|\hat{\bms}-\bms\|^2}$, which results in the widely-used linear minimum MSE (L-MMSE) equalization matrix
\begin{align} \label{eq:LMMSEmatrix}
\bW^\Herm = \textstyle \left(\widetilde{\bH}^\Herm\widetilde{\bH} + \frac{\No}{\Es}\bI_U\right)^{-1}\widetilde{\bH}^\Herm,
\end{align}
where $\widetilde{\bH}\in\complexset^{B'\times U}$ is the channel estimated during a training phase.
While common L-MMSE equalizer implementations use high-resolution numbers (typically 10-bit to 12-bit) for the entries of $\bW^\Herm$ (see, e.g., \cite{wu2014large}), it was shown in \cite{castaneda19fame} that the corresponding hardware can be power hungry in large-bandwidth applications, as it is the case for mmWave systems. 

In order to reduce the power consumption of the equalization step $\hat{\bms}=\bW^\Herm\bmz$, the work in \cite{castaneda19fame} proposes \emph{finite-alphabet equalization}, which uses carefully-designed low-resolution equalization matrices of the form $\bV^\Herm=\mathrm{diag}\left(\boldsymbol\mu\right)\bX^\Herm$. Here, the entries of the matrix $\bX^H\in\setX^{U\times B'}$ are taken from a reduced-precision alphabet $\setX$ (e.g., represented using 1-bit to 4-bit numbers). The vector $\boldsymbol\mu$ contains post-equalization scaling factors that are represented with more bits (e.g., 10-bit). 
In order to obtain unbiased estimates of the transmitted symbols, the entries of the scaling vector $\boldsymbol\mu$ are set to \cite{castaneda20soft}
\begin{align} \label{eq:posteqscaling}
    \mu_u=(\bmx^\Herm_u\tilde{\bmh}_u)^{-1}, \quad u=1,\ldots,U,
\end{align} 
and the resulting estimates are computed as follows: 
\begin{align} \label{eq:unbiasedEq}
\hat{s}_u = \bmv_u^H\bmz = \mu_u \bmx^\Herm_u \bmz, \quad u=1,\ldots,U.
\end{align}
Since the vectors $\bmx^\Herm_u$, $u=1,\ldots,U$, contain low-resolution entries, computing the inner products $\bmx^\Herm_u \bmz$ in \fref{eq:unbiasedEq} can be accomplished with low-resolution digital circuitry that occupies small area and consumes low power~\cite{castaneda19fame,castaneda20tcas}. Post-equalization scaling only needs to be calculated once per UE when computing the estimates $\hat{s}_u$, $u=1,\ldots,U$, and therefore does not dominate the equalizer's power consumption. 

In~\cite{castaneda19fame}, two methods are proposed to determine the entries of the low-resolution matrix $\bX^\Herm$, offering a trade-off between the complexity of calculating $\bX^H$ and the error-rate performance of the resulting finite-alphabet equalizer.
In this work, we will use the finite-alphabet L-MMSE (FL-MMSE) method, which has lower complexity but exhibits inferior performance for 1-bit and 2-bit resolutions.
As detailed in~\cite{castaneda19fame}, FL-MMSE computes the rows $\bmx_u$ of $\bX^\Herm$ by uniformly quantizing the rows $\bmw_u$ of the L-MMSE equalization matrix~$\bW^\Herm$ in \fref{eq:LMMSEmatrix}, according to $\bmx_u=Q_k(\Re\{\bmw_u\})+jQ_k(\Im\{\bmw_u\})$, where $Q_k(\cdot)$ is given by \fref{eq:quantizer} with $\Delta=\|\bmw_u\|_{\widetilde\infty} 2^{1-k}$.
Finally, the scaling factors~$\mu_u$, $u=1,\ldots,U$, are calculated from the vectors $\bmx^\Herm_u$ as in \fref{eq:posteqscaling}. 


\section{Power Models}
\label{sec:power}

We now introduce our model for the total power consumption~$P$ of the ADC array and the spatial equalizer given by
\begin{align} 
P=P_\text{ADC}+P_\text{EQ},
\end{align}
where $P_\text{ADC}$ and $P_\text{EQ}$ stand for the power consumption of the ADC array and the spatial equalizer, respectively. 
These two power models depend on the number of ADC bits~$q$, the number of FL-MMSE bits~$k$, the number of UEs $U$, the number of active BS antennas $B'$, and the ADC sampling rate~$f_s$. 

\subsection{Power Model for the ADC Array}
To model the power consumption $P_\text{ADC}$ of the ADC array consisting of $2B'$ active converters that sample the I/Q baseband signals, we use Walden's figure of merit ($\text{FoM}_\text{W}$)~\cite{panagiotis20} to obtain
\begin{align} \label{eq:adcpower}
P_\text{ADC} = \text{FoM}_\text{W} \times 2^q \times 2B'  \times f_\text{s}.
\end{align}
In what follows, we will use a $\text{FoM}_\text{W}$ of $70.8$\,fJ per conversion step, which was provided in~\cite{jiang21adc} for a 28\,nm CMOS successive-approximation register ADC with $f_\text{s}=2$\,GS/s.

\subsection{Power Model for the Finite-Alphabet Equalizer}
To model the power consumption $P_\text{EQ}$ of a finite-alphabet equalizer, we consider the 28\,nm CMOS design presented in~\cite{castaneda20tcas}, which supports a baseband sampling rate of $f_\text{s}=2$\,GS/s. 
This design consists of multiple time-interleaved instances of the Parallel Processor in Associative Content-addressable memory (PPAC)~\cite{castaneda2019ppac}, a spatial architecture that uses $4kB'U$ processing-in-memory bit-cells to compute a complex-valued matrix-vector product between $\bX^H$ and $\bmz$ in $q$ clock cycles. 

Since the equalization operation in \fref{eq:unbiasedEq} is independent for each UE, the power consumption is directly proportional to $U$.
While the number of PPAC bit-cells is directly proportional to~$k$ and $B'$, varying~$k$ and $B'$ also affects the power consumption of the logic in PPAC that processes the bit-cells' outputs.
Modeling the effect of $k$ and $B'$ on this surrounding logic is non-trivial.
However, since the bit-cells represent a significant component of the design, it is reasonable to use a linear model as in~\cite{xu17a, abdelghany18} for the total power of the equalizer.
As a matter of fact, the power consumption of the PPAC implementations in~\cite{castaneda20tcas} scales linearly with $k$.
We approximate the equalization power of PPAC as directly proportional to $B'$.
Finally, the throughput of PPAC is inversely proportional to the ADC resolution $q$.
Consequently, the number of time-interleaved PPAC instances required to reach the target $f_\text{s}=2$\,GS/s is directly proportional to $q$, just like $P_\text{EQ}$.
Putting it all together, we arrive at
\begin{equation} \label{eq:eqpower}
P_\text{EQ} = (2.44k-0.48) \times \frac{q}{7} \times \frac{B'}{256} \times \frac{U}{16} \times \frac{f_\text{s}}{2\,\text{GS/s}} \quad \text{[W]}.
\end{equation}
Here, the term $2.44k-0.48$ is obtained from a linear fit for the PPAC power results provided in \cite{castaneda20tcas}, which are for a system with $B=B'=256$, $U=16$, $q=7$, and $k\in\{1,2,3\}$.

We note that the model in \fref{eq:eqpower} only considers the power consumption for applying equalization as in $\hat{\bms}\!=\!\bV^\Herm \bmz$.
We do \emph{not} consider the power consumption for preprocessing, i.e., computing the L-MMSE matrix in \fref{eq:LMMSEmatrix} and extracting the FL-MMSE matrix.
While equalization must be carried out for every received vector at baseband sampling rate, preprocessing is required only once per channel coherence interval, which can last for more than $10^4$ samples in  mmWave systems \cite{Bogale15}.


\section{Results}
\label{sec:simres}

\begin{figure*}[tp]
\centering
\subfigure[mmMAGIC LoS, $16$-QAM]{\includegraphics[width=.65\columnwidth]{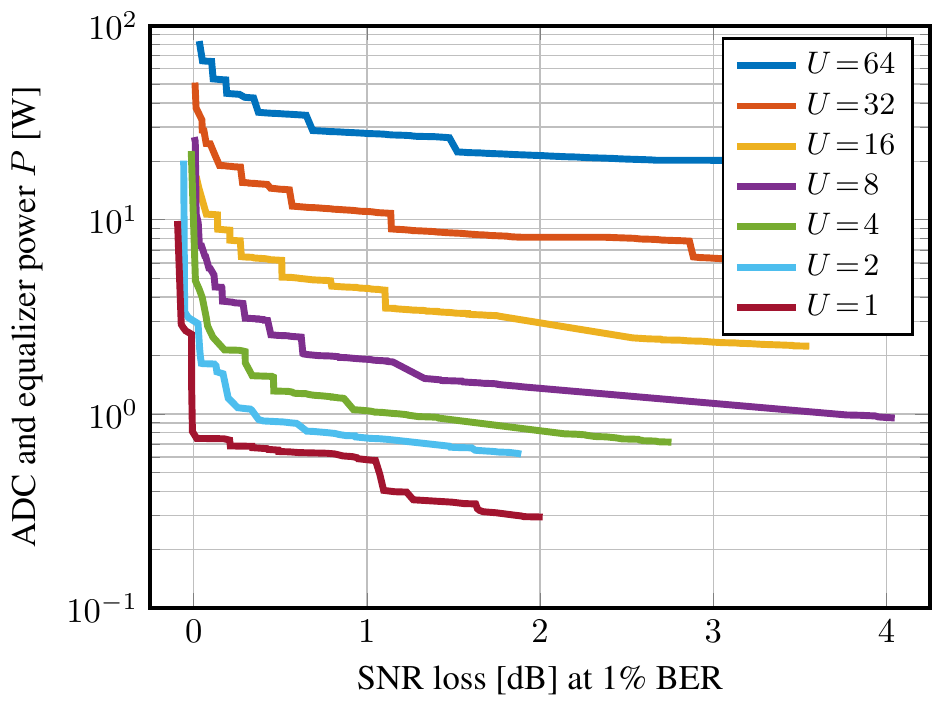}\label{fig:power_los_16qam}}
\hfill
\subfigure[mmMAGIC LoS, QPSK]{\includegraphics[width=.65\columnwidth]{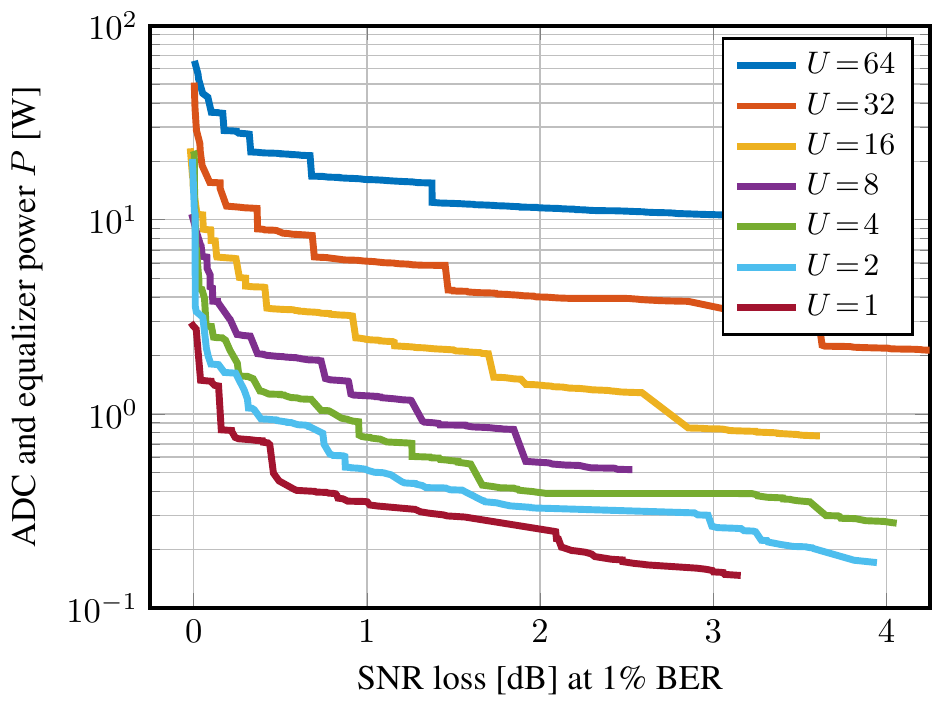}\label{fig:power_los_qpsk}}
\hfill
\subfigure[mmMAGIC non-LoS, QPSK]{\includegraphics[width=.65\columnwidth]{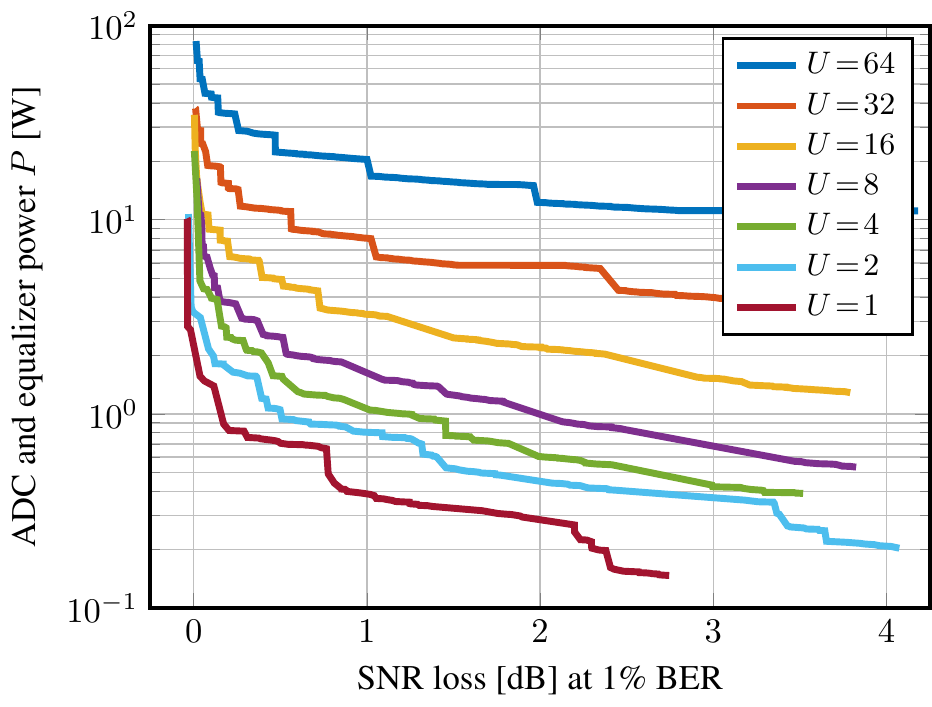}\label{fig:power_nlos_qpsk}}
\caption{ADC array and spatial equalizer power consumption $P=P_\text{ADC}+P_\text{EQ}$ for Pareto-optimal configurations in a $B=256$ BS antenna system. For each UE load $U$, we vary the number of ADC bits $q$, equalization bits $k$, and active BS antennas $B'$. Our results for both LoS and non-LoS channels with QPSK and $16$-QAM communication reveal that the power consumption can be reduced by two orders of magnitude depending on the communication scenario.}\label{fig:power_all}
\end{figure*}

\begin{figure*}[tp]
\centering
\subfigure[LoS, $0.1$\,dB SNR loss, $16$-QAM]{\includegraphics[width=.65\columnwidth]{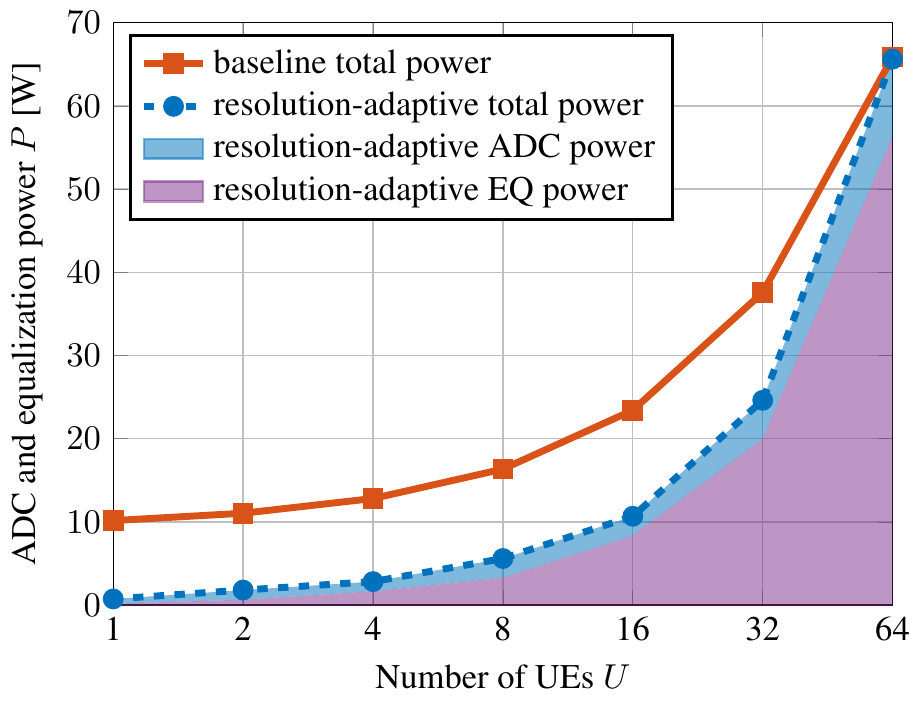}\label{fig:comp_baseline}}
\hfill
\subfigure[LoS, $0.1$\,dB SNR loss, QPSK]{\includegraphics[width=.65\columnwidth]{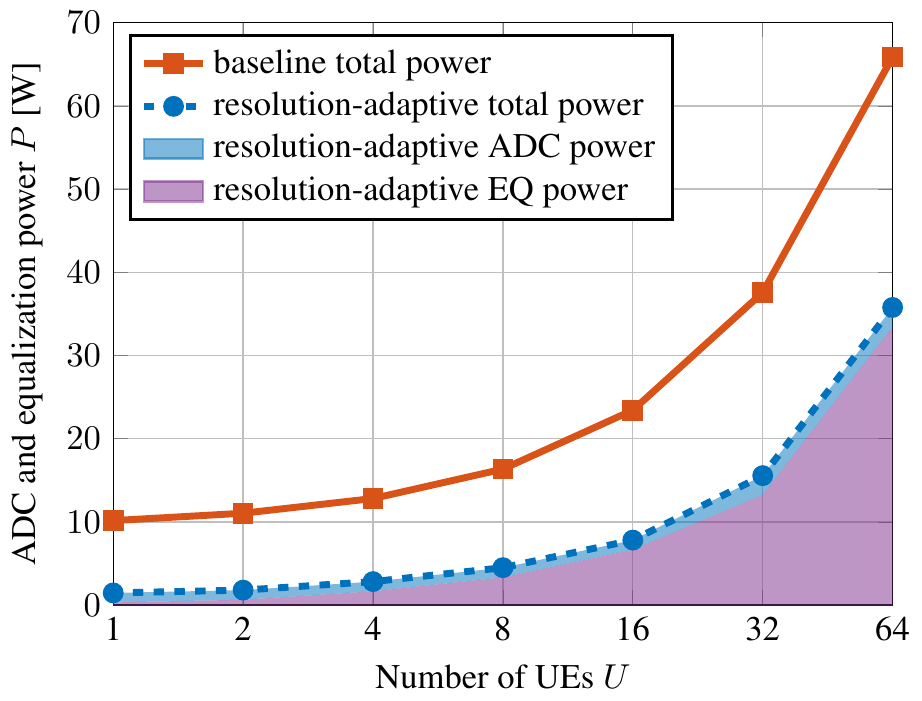}\label{fig:comp_mod}}
\hfill
\subfigure[LoS, $0.5$\,dB SNR loss, QPSK]{\includegraphics[width=.65\columnwidth]{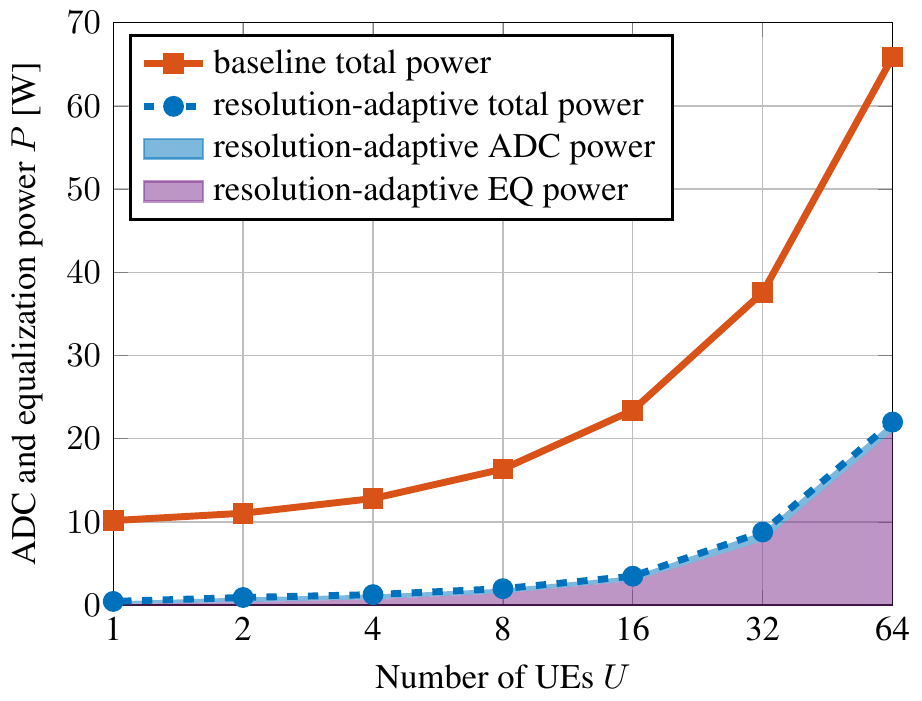}\label{fig:comp_snr_tradeoff}}
\caption{ADC array and spatial equalizer power consumption $P=P_\text{ADC}+P_\text{EQ}$ for Pareto-optimal configurations as the number of UEs $U$ varies. The baseline corresponds to a system  designed for the worst case of $U=64$ with $16$-QAM that is only able to adapt the equalizer's active processing elements to the UE load~$U$. Resolution-adaptive designs can achieve up to $22\times$ power savings, depending on the UE load~$U$, modulation scheme, and SNR loss at $1$\% BER.}\label{fig:comp_all}
\end{figure*}

\subsection{Simulation Setup and Performance Metrics}
We consider a mmWave massive MU-MIMO uplink system in which a BS with a $B=256$ antenna $\lambda/2$-spacing uniform linear array (ULA) is receiving data from $U\in\{1,2,4,8,16,32,64\}$ UEs transmitting at $f_\text{s}=2\,$GS/s and a carrier frequency of $60\,$GHz.
We perform Monte-Carlo simulations using line-of-sight (LoS) and non-LoS channel matrices generated with QuaDRiGA \cite{jaeckel2014quadriga} using the mmMAGIC UMi scenario. We model situations in which the UEs are randomly placed at distances ranging from $10\,$m to $100\,$m in front of the BS, within a $120^\circ$ sector, and a minimum angular separation of $1^\circ$.
We furthermore assume $\pm3\,$dB per-UE power control as described in \fref{sec:system}.
We estimate the channel using a pilot-based least squares (LS) estimator followed by denoising with  beamspace channel estimation (BEACHES)~\cite{mirfarshbafan19a}.

In order to evaluate the effect that the number of ADC bits~$q$, finite-alphabet equalization bits $k$, and active BS antennas $B'$ have on the system's error-rate performance and power consumption, we vary $q\in\{1,2,\ldots,8,\infty\}$, $k\in\{1,2,\ldots,6,\infty\}$, and $B'\in\{232,233,\ldots,256\}$. 
A value of $q=\infty$ corresponds to bypassing the ADCs, i.e., $\bmz=\hat{\bmy}$; setting $k=\infty$ corresponds to unbiased L-MMSE equalization without quantization. 
For each configuration of $q$, $k$, $B'$, and $U$, we quantify the error-rate performance by finding the minimum SNR at which $1\%$ uncoded bit-error rate (BER) is attained, and compute the SNR loss (in decibels) with respect to the ideal case in which $q=k=\infty$ and $B'=B=256$.
For each configuration of $q$, $k$, $B'$, and $U$, we also determine the resulting power consumption $P=P_\text{ADC}+P_\text{EQ}$ using the models in \fref{eq:adcpower} and \fref{eq:eqpower}.
For all of the considered configurations with a fixed number of UEs $U$, we find the Pareto-optimal envelope of the designs that achieve the lowest power consumption under the same SNR loss, or that achieve the lowest SNR loss at the same power consumption, i.e., the lower envelope of the (SNR loss, power $P$) data points generated from all considered $(q,k,B')$ configurations.
The results are shown in \fref{fig:power_all} and discussed next. 

\subsection{SNR Operating Point vs.\ Power Trade-Offs}
\fref{fig:power_los_16qam} demonstrates that the proposed resolution-adaptive architecture enables one to trade-off performance (measured in terms of the SNR loss) versus the ADC and equalizer power~$P$.
For a fixed UE load $U$, we see that by allowing a larger SNR loss, the BS power can be reduced considerably. 
Indeed, by lowering the ADC and equalizer resolutions $q$ and $k$, respectively, as well as the number of active BS antennas $B'$, power consumption is reduced, whilst requiring a higher SNR to achieve the same target BER of $1$\%.
This implies that for situations in which a larger SNR loss can be tolerated, one can drastically reduce the system's power consumption.
Consider, for example, \fref{fig:power_los_16qam} with $U=8$ UEs. For an SNR loss of $0$\dB, the power consumption would be $P=26.6$\,W whereas for an SNR loss of $1.5$\,dB, the power consumption is only $P=1.5$\,W. 
Hence, such resolution-adaptive architectures can reduce the power consumption by one order of magnitude for a given UE load.

When comparing \fref{fig:power_los_16qam} with \fref{fig:power_los_qpsk}, we furthermore observe that equalization of QPSK symbols instead of $16$-QAM can further reduce the power consumption, which implies that simpler equalization problems require less work.
Furthermore, when comparing \fref{fig:power_los_qpsk} with \fref{fig:power_nlos_qpsk}, we can see that the same performance and power trade-off persists under LoS and non-LoS propagation conditions, respectively.
This indicates that resolution-adaptive architectures provide similar gains for different mmWave propagation~scenarios.

\subsection{Communication Scenario vs.\ Power Trade-Offs}
\fref{fig:power_all} also illustrates how the total power consumption~$P$ spans a wide range as the load factor, modulation scheme, and propagation conditions vary.
Hence, fully exploiting all of these dynamic aspects in the BS' power consumption requires a \textit{resolution-adaptive} architecture (cf.~\fref{fig:system_overview}) that is able to dynamically adapt to these  factors.

In order to further demonstrate the potential of such a resolution-adaptive BS architecture, we perform the following experiment.
First, we consider a baseline BS architecture with $B'=B=256$ (active) antennas that can adapt \emph{only} its equalizer to the UE load $U$ as follows: Since linear equalization is an independent problem per UE, a linear equalizer architecture can easily be adapted so that the equalizer power $P_\text{EQ}$ is directly proportional to $U$.
However, the number of ADC bits $q$ and equalization bits $k$ for this baseline architecture is fixed so that the worst SNR loss experienced across all different considered scenarios (load factors, modulation schemes, and propagation conditions) is below $0.1$\,dB. In our simulations, this worst-case scenario is $U=64$ with $16$-QAM (under both LoS and non-LoS propagation conditions), and requires $q=7$ and $k=6$.
The power consumption $P$ of such a baseline architecture is shown with a solid line in \fref{fig:comp_all}.

In contrast, the proposed resolution-adaptive architecture is not only able to optimize its equalization power $P_\text{EQ}$ as a function of the number of UEs $U$, but it can also dynamically adapt the ADC resolution $q$, equalization resolution $k$, and the number of active antennas $B'$. 
The power consumption of such resolution-adaptive architecture is shown with a dashed line in \fref{fig:comp_baseline} when operating with $16$-QAM under LoS propagation.
Clearly, a resolution-adaptive architecture has to support the same worst-case scenario as the baseline and, hence, the power consumption is the same for $U\!=\!64$ UEs.
However, once we consider smaller load factors $\beta$, the resolution-adaptive architecture reduces the power consumption by $2.2\times$ for $U\!=\!16$ and reaches maximum power savings of $14\times$ at $U\!=\!1$. 
In \fref{fig:comp_mod}, the modulation scheme changes from $16$-QAM to QPSK. The baseline architecture cannot exploit this change, whereas a resolution-adaptive design can readjust $q$, $k$, and $B'$ to reduce power consumption by at least $1.8\times$ when $U\!=\!64$.
Furthermore, \fref{fig:comp_snr_tradeoff} illustrates how further power savings can be achieved by allowing a larger SNR loss of $0.5\,$\dB, significantly reducing power by $6.7\times$ when $U\!=\!16$ and up to a factor of $22\times$ when $U\!=\!1$.
Finally,  the shaded areas in \fref{fig:comp_all} show how much power is consumed by the ADC array and equalizer of the resolution-adaptive design. We observe that the equalizer dominates the total power consumption, except for very small load factors of~$U\leq2$.


\section{Conclusions}
We have proposed a resolution-adaptive ADC array and all-digital spatial equalization architecture for mmWave massive MU-MIMO, which is able to optimize the system's power consumption depending on the load factor, modulation scheme, and channel conditions. 
By combining low-resolution ADCs with finite-alphabet equalization, the power and performance (measured as the minimum SNR required to achieve a target BER) of all-digital architectures is tunable with respect to the ADC bits $q$, equalization bits $k$, and number of active BS antennas $B'$.
We have shown that resolution adaptivity enables a wide trade-off region between power consumption and performance, with a power consumption ranging from $83$\,W for $U=64$ UEs to only $0.15$\,W  for $U=1$ UE in a $256$-antenna mmWave massive MU-MIMO BS supporting a bandwidth of $1$\,GHz.
In summary, our results demonstrate that resolution adaptivity is a critical component of power-efficient all-digital mmWave massive MU-MIMO BS designs. 

We see the following avenues for future work. Augmenting our analysis with the impact on power and performance of RF circuitry (including low-noise amplifier, mixer, etc.) would provide a more comprehensive assessment of the benefits of resolution-adaptive BS designs. A theoretical analysis of the trade-offs shown in this paper is part of ongoing~work.

\balance

\balance

\end{document}